\documentclass[reprint,twocolumn,aps,prb,showpacs]{revtex4-2}

\usepackage{amsmath}
\usepackage{amssymb}
\usepackage{graphicx}
\usepackage{dcolumn}
\usepackage{bm}
\usepackage[colorlinks=true, allcolors=blue]{hyperref}
\usepackage{epstopdf}
\usepackage[utf8]{inputenc}
\usepackage{amssymb}
\usepackage[dvipsnames]{xcolor}

\begin{document}

\title{Quantitative determination of the orbital-selective Mott transition and quantum entanglement in the orbital-selective Mott phase}

\author{Yuekun Niu$^{1}$}
\email[]{ykniu@imu.edu.cn}%

\author{Yu Ni$^{2}$}

\author{Haishan Zhang$^{1}$}

\author{Liang Qiu$^{3}$}

\author{Jianli Wang$^{3}$}

\author{Leiming Chen$^{3}$}



\author{Yun Song$^{4}$}

\email[]{yunsong@bnu.edu.cn}%

\author{Shiping Feng$^{4}$}

\affiliation{$^{1}$School of Physical Science and Technology, Inner Mongolia University,
Hohhot 010021, China}

\affiliation{$^{2}$College of Physics and Electronic Information, Yunnan Normal University,
Kunming 650500, China}

\affiliation{$^{3}$School of Materials Science and Physics, China University of Mining and
Technology, Xuzhou 221116, China}

\affiliation{$^{4}$Department of Physics, Beijing Normal University, Beijing 100875, China}

\date{\today}
	
\begin{abstract}
We examine the orbital-selective Mott transition in the non-hybridized two-band Hubbard model using the dynamical mean-field theory.
We find that the orbital-selective Mott transition could be quantitatively depicted by the {local two-qubit fidelity}. 
Furthermore, within the orbital-selective Mott phase, the combined characteristics of the two orbitals lead to the presence of quantum entanglement, which is characterized by the non-semi-integer values of local two-qubit fidelity.
It is demonstrated that the Hund's coupling results in the ground states of both wide and narrow bands exhibiting the specific superposition states, indicating the existence of quantum entanglement within orbital-selective Mott phase.
Without Hund's coupling, there are no specific superposition states, nor does quantum entanglement occur within the orbital-selective Mott phase.
The mechanisms underlying the orbital-selective Mott transition show prominent variations depending on the presence or absence of Hund's coupling and its transverse terms.

\end{abstract}

\maketitle

\section{Introduction} 
The electron correlation and orbital degeneracy give rise to a variety
of intriguing phenomena near the metal-insulator transition (MIT) in strongly correlated systems \cite{Imada1998,Kotliar2006,Rohringer2018}. In a multiorbital system, the competing energy scales associated with the Coulomb interaction and orbital differentiation lead to the orbital-selective Mott transition (OSMT), where the carriers in a subset of orbitals become localized while the others remain metallic \cite{Anisimov2002,Koga2004}.
It has been demonstrated that the microscopic origin of OSMT can be attributed to the spin-flip and pair-hopping processes of Hund's coupling \cite{Liebsch2005,Medici2011,Georges2013,Yu2013,Rinc2014}.
Furthermore, the difference in bandwidths between different bands \cite{Arita2005,Inaba2006,Medici2005,Lee2010,Wu2021} and the presence of crystal field splitting \cite{Werner2007,Medici2009,Song2009,Kugler2019} have also been verified to contribute significantly to the OSMT phenomenon.
Currently, orbital-selective behavior is a very active field producing new surprises and can contribute to many novel and interesting quantum effects \cite{Khomskii2021}.

Quantum entanglement is assumed to be a key resource used to perform interesting physical tasks \cite{Osborne2002}.
It is considered an important quantity for understanding of correlations, transport properties, and phase transitions in many-body systems \cite{Luigi2008,Tichy2011}.
The properties of entanglement in many-body systems have become an important part of quantum information and condensed matter physics \cite{Luigi2008}.
One of the fundamental questions in entanglement theory concerns the quantification of entanglement \cite{Horodecki2009}.
Entanglement entropy (von Neumann entropy) is a commonly used method to measure entanglement \cite{Chen2022,Dana2020,Lev2017}.
For instance, Gu {\it et al}. \cite{Gu2004} demonstrated the relation between entanglement and quantum phase transition, showing that entanglement can be used to identify quantum phase transitions in fermionic systems.
Electrons, as potential carriers of entanglement, can carry entanglement in their spin and spatial degrees of freedom, with entanglement resulting from interactions \cite{Tichy2011}.
A question arises regarding the close relation between quantum entanglement and the OSMP.
Therefore, possessing a tool capable of verifying the manifestation of entanglement behavior in strongly correlated systems holds immense value.

Recently, the discovery of the interband doublon-holon bound state  \cite{N2018,Niu2019,Kugler2019,Hallberg2020,Aucar2021}
has indicated the coexistence of a single-hole state in one band and a doubly-occupied state in another band within the orbital-selective Mott phase (OSMP).
This finding highlights the inherent correlation of different orbitals, emphasizing that their behavior cannot be regarded as independent within the OSMP.
Additionally, the local quantum state fidelity (LQSF) \cite{Niu2023} provides a convenient approach to determining the critical point of the Mott transition for the one-band Hubbard model.
Notably, the ground-state wave function in LQSF includes four occupied states (zero, spin-up, spin-down, and double-occupied states) of the single impurity site.
These four occupied states can be regarded as the computational ground state of two qubits \cite{Nielsen2010}.
Multiple qubits allow for the implementation of second-generation quantum techniques, including entanglement and state squeezing, which provide a true quantum advantage that cannot be realized with classical sensors \cite{Degen2017}. 
In the present paper, we extend LQSF to the half-filled two-band Hubbard model.

The main purpose of this paper is to quantitatively describe OSMT and directly measure the behavior of quantum entanglement in multiorbital systems.
We study the non-hybridized two-band Hubbard model using dynamical mean-field theory (DMFT) with the Lanczos algorithm \cite{Dagotto1994}.
The continuity properties of local entanglement \cite{Gu2004} are investigated to confirm the existence of quantum entanglement \cite{Chen2022,Dana2020,Lev2017} in a one-dimensional lattice model. 
In this paper, we extend it to the half-filled two-band Hubbard model as an auxiliary tool to indirectly demonstrate interorbital entanglement.
We examine quantum entanglement and quantitatively determine the phase transition points in the two-band Hubbard model, providing a perspective on the Mott MIT of strongly correlated multiorbital systems.

The paper is organized as follows. In Sec. \ref{sec:mm}, we introduce the theoretical model and the DMFT numerical methodology.
Section \ref{sec:rs} presents a comprehensive overview of the results, encompassing the characterization of OSMT and quantum entanglement through local two-qubit fidelity, the exploration of Hund’s coupling driven quantum entanglement, and the analysis of the roles of longitudinal and transverse Hund’s couplings on quantum entanglement.
The main conclusions are summarized in Sec. \ref{sec:cs}.

\section{Models and methods\label{sec:mm}} 
The Hamiltonian of the two-band Hubbard model \cite{Roth1966,Oles1983} is given by
\begin{eqnarray}\label{hubbard}
\hat{H}=&&-\sum_{\langle i j \rangle l\sigma}t_{l}d_{i l\sigma}^\dag d_{j l\sigma}
-\mu\sum_{i l\sigma}d_{i l\sigma}^\dag d_{i l\sigma}\nonumber\\
&&+\frac{U}{2}\sum_{i l\sigma}n_{i l\sigma}n_{i l\bar{\sigma}}
+ \sum_{i\sigma\sigma^{\prime}}(U^{\prime} - \delta_{\sigma\sigma^{\prime}}J_{z})n_{i1\sigma}
n_{i2\sigma^{\prime}}	\nonumber\\
&&+\frac{J_{ph}}{2}\sum_{i,l\neq l^{\prime},\sigma}d_{i l\sigma}^\dag
d_{i l\bar{\sigma}}^{\dag}d_{i l^{\prime}\bar{\sigma}}d_{i l^{\prime}\sigma}
\nonumber\\
&&+\frac{J_{sf}}{2}\sum_{i,l\neq l^{\prime},\sigma\sigma^{\prime}}d_{i l\sigma}^\dag
d_{i l^{\prime}\sigma^{\prime}}^{\dag}d_{i l\sigma^{\prime}}d_{i l^{\prime}\sigma},
\end{eqnarray}
where the summation $\langle ij \rangle$ is for each site $i$, restricted to its nearest-neighbor (NN) sites $j$, $d_{i l\sigma}^\dag(d_{i l\sigma})$ is the creation (annihilation) operator for an electron in orbital $l (=1$ or $2)$ with spin $\sigma$ of
lattice site $i$, and $n_{i l\sigma}$ is the occupation number operator of electrons in orbital $l$ of lattice site $i$.
$t_{l}$ denotes the NN orbital hopping amplitude in orbital $l$, $\mu$ is the chemical potential, and $U$ ($U^{\prime}$) represent the intraorbital (interorbital) Coulomb interactions.
The Hund's couplings consist of the Ising-type term $J_{z}$, spin-flip term $J_{sf}$, and pair-hopping term $J_{ph}$. Here we set $J_{z}=J_{sf}=J_{ph}=J$.
For the system with spin rotation symmetry, the relationship $U=U^{\prime}+2J$ should be kept \cite{Castellani1978,Raymond1997}.
The hopping amplitude ratio, also known as the bandwidth ratio, is defined as $R=t_{2}/t_{1}$, where $t_{1}$ represents the NN hopping amplitude of the wide band, and is set to 1 unless explicitly stated.

In the framework of DMFT \cite{Georges1996}, the two-band Hubbard model is mapped onto an
effective single impurity model,
\begin{eqnarray}\label{anderson}
\hat{H}_{imp}=&&\sum_{m l\sigma}\varepsilon_{m l}c_{m l\sigma}^\dag c_{m l\sigma}
+\sum_{m l\sigma}V_{m l}(c_{m l\sigma}^\dag d_{l\sigma}+d_{l\sigma}^\dag c_{m l\sigma})
\nonumber\\
-&&\mu\sum_{l\sigma}d_{l\sigma}^\dag d_{l\sigma}+\hat{H}_{int}[d_{l\sigma}],
\end{eqnarray}
where $d_{l\sigma}^{\dag}(d_{l\sigma})$ creates (annihilates) an electron in the
{\it impurity site} for orbital $l$, and $c_{m l\sigma}^{\dag}(c_{m l\sigma})$ creates
(annihilates) an electron in a {\it conduction band} for orbital $l$.
The {\it impurity site} and {\it conduction band} are coupled to each other via effective
parameters $\varepsilon_{m l}$ and $V_{m l}$. These parameters are determined by
performing the self-consistent DMFT calculations, and the mapping becomes exact in the
limit of infinitely lattice coordination \cite{Georges1992}.
We introduce the local electron Green's function in real-space as \cite{Anisimov2010,Mahan2000}
$\label{local-electron-Green-function}
\mathcal{G}_{\sigma}(\tau)=-<T_{\tau}d_{\sigma}(\tau)d_{\sigma}^{\dag}(0)>$
with the imaginary time $\tau=it$.
This local electron Green's function in energy-space can be obtained directly by performing the Fourier transformation.
\begin{eqnarray}
\mathcal{G}_{\sigma}(i\omega_{n})&=&\int_{0}^{\beta}d\tau e^{i\omega_{n}\tau}
\mathcal{G}_{\sigma}(\tau),\\
\mathcal{G}_{\sigma}(\tau)&=&\frac{1}{\beta}\sum_{n=-\infty}^{\infty}e^{-i\omega_{n}\tau}
\mathcal{G}_{\sigma}(i\omega_{n}),\label{gg}
\end{eqnarray}
where $-\beta\leq\tau\leq\beta$ and the fermionic Matsubara frequency $\omega_{n}=(2n+1)\pi/\beta$ with $n=0,\pm1,\pm2,\cdots$.
The local Green's function on a Bethe lattice with a semicircular density of states can be obtained
via a single-site impurity problem supplemented by the following self-consistency
relation \cite{caffarel1994,Laloux1994},
$\mathcal{G}_{0l\sigma}^{-1}(i\omega_{n})=i\omega_{n}
+\mu-t_{l}^{2}\mathcal{G}_{l\sigma}(i\omega_{n})$, where $\mathcal{G}_{0}$ is the bare Green's function. 
The self-consistency relation ensures that the on-site (local) component of the Green's function [$\mathcal{G}_{ii}(i\omega_{n})=\sum_{k}\mathcal{G}(k,i\omega_{n})$] coincides with the Green's function $\mathcal{G}(i\omega_{n}$) calculated from the effective action.

In a recent paper \cite{Niu2023}, it has been demonstrated that LQSF, as a proper physical quantity,
can provide a convenient approach to determining the critical points of MIT in the one-band Hubbard model.
LQSF was proposed through a method of analogy to the symmetry-protected topological order (SPT) \cite{Chen2011,Lan2017} and the concept of quantum fidelity \cite{Zhou2008,Rams2011}.
Initially, it was observed that gapped quantum states exhibit short-range entanglement, which corresponds to a SPT order as described in Ref. \cite{Lan2017}.
We extend the classification method of the SPT phases in higher dimensions to label gapped quantum phases based on the four occupation states of electrons on an impurity site according to Ref. \cite{Chen2011}.  
Furthermore, the quantum fidelity is purely a quantum information concept usually defined as the overlap between two quantum states, while quantum phase transitions are intuitively accompanied by an abrupt change in the structure of the ground-state wave function. 
Particularly in the proximity of critical regions, slight variations of the Hamiltonian parameters can lead to significant alterations in the ground state caused by the distinct structures in different phases.
Therefore, we introduced the ground-state fluctuations of spin occupation on a single impurity site to denote the LQSF,
which can serve as a tool to measure the evolution of the ground state with the Coulomb interaction $U$.
In our formulation, the ground-state wave function is decomposed into a superposition of spin-up, spin-down, zero, and double occupied states, where $|\Phi_{imp}^{o}\rangle$ can be regarded as a two-qubit state \cite{Nielsen2010}.
This is the essential difference between LQSF and quantum fidelity.
To avoid confusion among readers because of the names of physical quantities, 
here we extend this method to multiorbital systems and rename it as the {\it local two-qubit fidelity }(LTQF),
\begin{equation}\label{Io}
L_{{\rm o}l}=-\frac{1}{\beta}\sum_{n=-\infty}^{\infty}e^{i\omega_{n}0^{+}}
\mathcal{G}_{l}(i\omega_{n})\langle \Phi_{imp}^{ol}|\hat{P}|\Phi_{imp}^{o^{\prime}l}\rangle,
\end{equation}
with $|\Phi_{imp}^{ol}\rangle$ given by
\begin{equation}
|\Phi_{imp}^{ol}\rangle=\sum_{s=1}^{4}p_{ls}|p_{ls}\rangle
=p_{l1}|0\rangle+p_{l2}|\uparrow\rangle+p_{l3}|\downarrow\rangle
+p_{l4}|\uparrow\downarrow\rangle,\nonumber
\end{equation}
where $\hat{P}$ is the net spin projection operator for the impurity site with $\langle0|\hat{P}|0\rangle=\langle\uparrow\downarrow|\hat{P}|\uparrow\downarrow\rangle=0$, $\langle\uparrow|\hat{P}|\uparrow\rangle=1$, and
$\langle\downarrow|\hat{P}|\downarrow\rangle=-1$.
$|\Phi_{imp}^{ol}\rangle$ ($|\Phi_{imp}^{o^{\prime}l}\rangle$) represents the ground-state wave function
of the single impurity site with an interaction strength of $U$ ($U+0^{+}$).
The ground state is represented by the superposition of spin-up, spin-down, zero, and double-occupied states.
Please note that our research is limited to temperatures of zero, and $\beta$ only serves as a frequency cutoff \cite{caffarel1994}.

\begin{figure}
\centering
\includegraphics[width=0.46\textwidth]{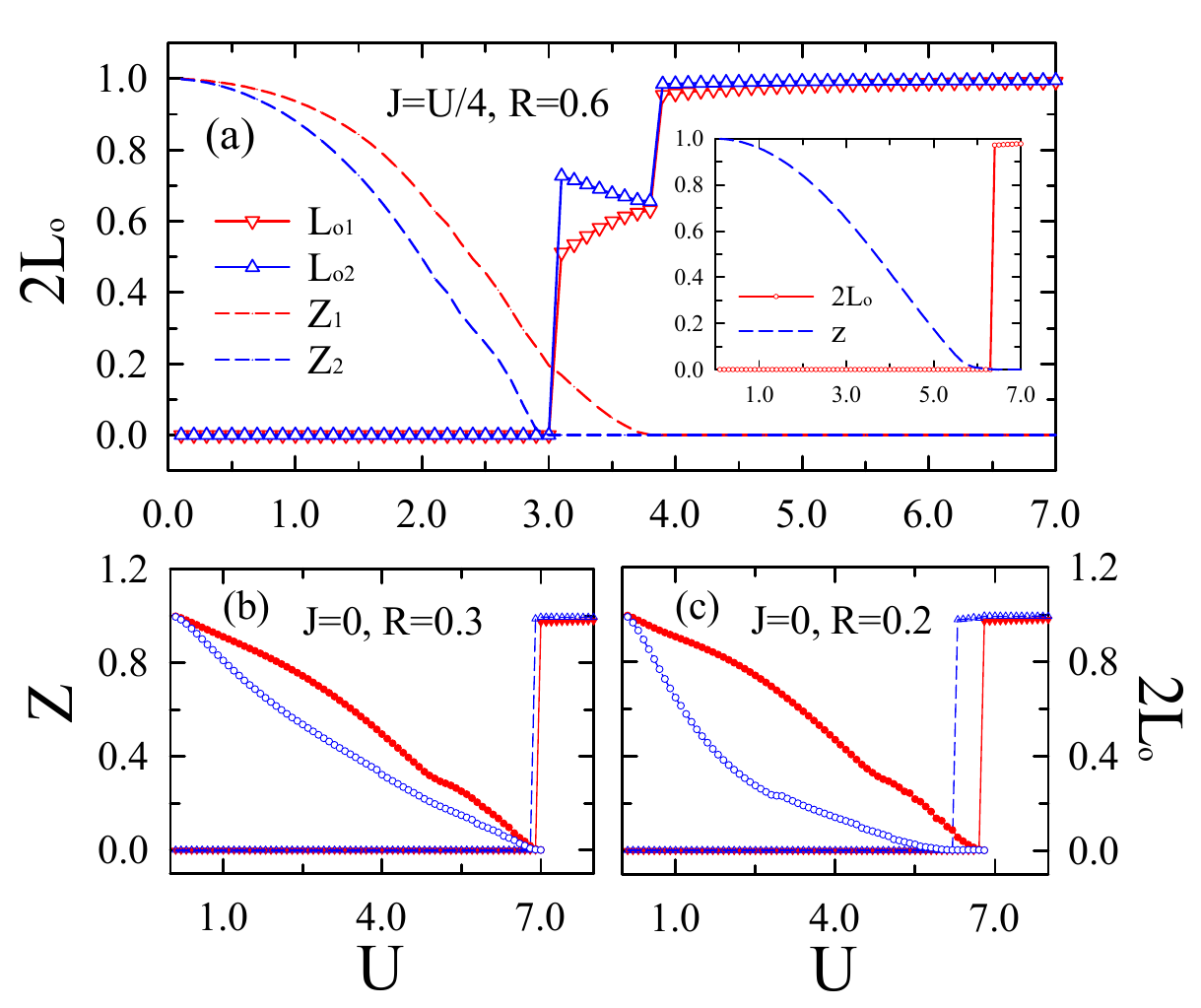}
\caption{(Color online) LTQF, denoted as $L_{\rm o1}$, $L_{\rm o2}$ and the quasiparticle coherent weights, represented by $Z_1$ and $Z_2$ for the wide and narrow orbitals, respectively, as a function of the Coulomb interaction $U$,
(a) $J=U/4, R=t_{2}/t_{1}=0.6$.
The Mott critical points are at $U_{c1}=3.8$ and $U_{c2}=3.0$ for the wide and narrow orbitals, respectively.
(Inset) The corresponding results of $L_{\rm o}$ and $Z$ of the one-band Hubbard model taken from Ref.~[\onlinecite{Niu2023}].
(b) $R=0.3, J=0$ $(U^{\prime}=U)$.   
(c) $R=0.2, J=0$ $(U^{\prime}=U)$. Closed (open) circles represent the $Z_1$ ($Z_2$), Triangle down (up) represent the $L_{\rm o1}$ ($L_{\rm o2}$).}
\label{fig:zl}
\end{figure}

\section{Results\label{sec:rs}}
\subsection{OSMT is quantitatively depicted by the LTQF}
We have conducted a series of calculations to investigate the OSMT in the two-band Hubbard model.
For example, we set the Hund's coupling as $J=U/4$ and the ratio of bandwidths as $R=t_{2}/t_{1}=0.6$.
LTQF is denoted as $L_{\rm o1}$ and $L_{\rm o2}$ for orbitals 1 (wide) and 2 (narrow) respectively,
along with the corresponding quasiparticle coherent weights $Z_1$ and $Z_2$, as a function of $U$, are plotted in Fig.~\ref{fig:zl}(a).
For comparison, the $L_{\rm o}$ and $Z$ of the one-band Hubbard model are also shown in Fig.~\ref{fig:zl}(a)(inset).
LTQF takes $L_{\rm o}=0$ for the metallic phase and $L_{\rm o}\approx 0.5$ for the insulating phase \cite{Niu2023}.
Moreover, we observe discontinuous jumps in the evolution of $L_{\rm o1}$ and $L_{\rm o2}$ with respect to $U$, which corroborate the occurrence of OSMT.
These results are consistent with the critical values of $U_{c1}\approx3.8$ and $U_{c2}\approx3.0$ obtained from the corresponding $Z_1$ and $Z_2$.

To highlight the advantage of LTQF in quantitatively describing the Mott transition, we investigated the impact of different bandwidth ratio $R$ values on OSMT when $J=0$ $(U^{\prime}=U)$. 
We find that there is no occurrence of OSMT as long as $R>0.3$, whereas OSMT exists when $R\leq0.3$ by employing LQSF as a means to describe the Mott transition.
As shown in Fig.~\ref{fig:zl}(b), LTQF changes suddenly at different critical points, $U_{c1}=6.9$ and $U_{c2}=7.0$ for $R=0.3$, which means the existence of OSMT.
As $R$ decreases, the difference between the two critical points becomes more pronounced, as shown in Fig.~\ref{fig:zl}(c) for $R=0.2$.
In comparison, the quasiparticle coherent weights of wide and narrow bands converge near the Mott transition, exhibiting a continuous decrease until they simultaneously reach zero at the critical point.
That is, the two bands undergo a common Mott transition at a single value of $U_{c}$. 
This result is in agreement with the conclusion that there is an OSMT for $R=0.15$ and no OSMT for $R=0.25$ at $J=0$ obtained in Ref. \cite{Medici2005}. 
However, it is inconsistent with the conclusion that there is no OSMT at $J=0$ by Liebsch \cite{Liebsch2003} and Koga \cite{Koga2004}. 

Note that the LTQF has been able to clearly distinguish the Mott critical points with different bands, but these critical points still cannot be distinguished using the quasiparticle coherent weights when $R=0.3$ [shown in Fig.~\ref{fig:zl}(b)]. 
Even at $R=0.2$, the quasiparticle coherent weights still show a zero at the same critical point when the system approaches the Mott transition [shown in Fig.~\ref{fig:zl}(c)].
As a result, regardless of whether it is the wide band (WB) or the narrow band (NB), it is easy to accurately determine the Mott critical points using LTQF.
This also solves the problem of difficulty in precisely determining the second critical point in Ref. \cite{Koga2004}.

\subsection{Quantum entanglement is directly measured by the LTQF}
\begin{figure}
\centering
\includegraphics[width=0.46\textwidth]{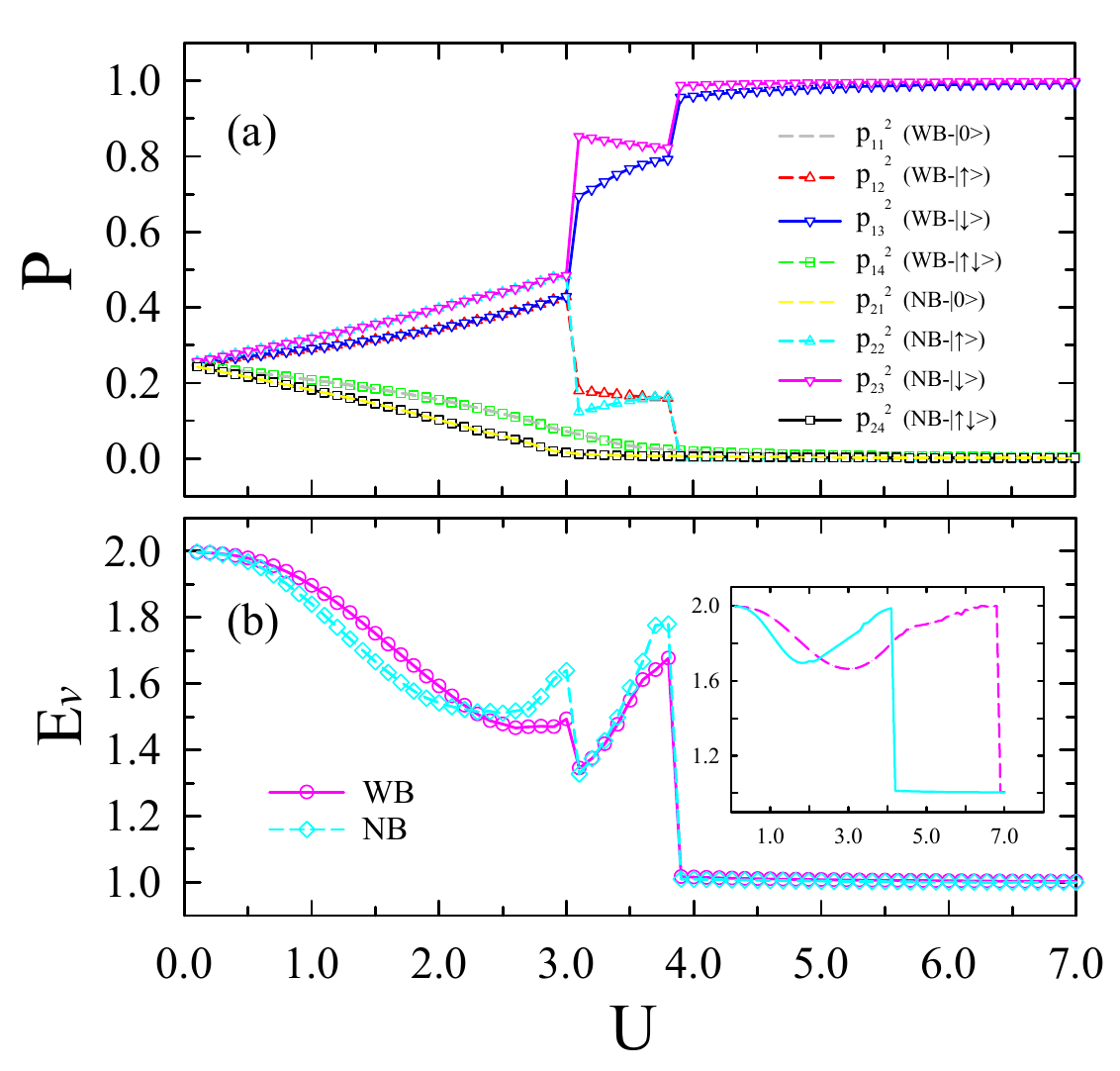}
\caption{(Color online) Interorbital entanglement within OSMP.
(a) The $U$ dependencies of the probabilities $p_{ls}^{2}$ of the zero-occupied state
$|0\rangle$, the spin-up occupied state
$|\uparrow\rangle$, the spin-down occupied state
$|\downarrow\rangle$, and double-occupied state
$|\uparrow\downarrow\rangle$ at the single impurity site.  
(b) The evolution of local entanglement $E_{v}$ of the wide band and narrow band with the $U$.
The model parameters are set to $J=U/4$ and $R=0.6$.
(Inset) The corresponding evolution results of $E_{v}$ in OSMP without interorbital entanglement using a single band model for $t_{1}=1.0$ and $t_{2}=0.6$}.
\label{fig:pev}
\end{figure}

In OSMP under the influence of a finite $J$, the behavior of LTQF unexpectedly deviates, exhibiting non-semi-integer values.
We attribute this phenomenon to the superposition state of electrons caused by quantum entanglement within OSMP, which is generated by the Hund's coupling. 
This conclusion is supported by the following factual evidence.
The superposition states of an impurity site can be expressed as $|\Phi_{imp}^{\rm o1}\rangle=p_{11}|0\rangle
+p_{12}|\uparrow\rangle+p_{13}|\downarrow\rangle+p_{14}|\uparrow\downarrow\rangle$ and
$|\Phi_{imp}^{\rm o2}\rangle=p_{21}|0\rangle+p_{22}|\uparrow\rangle+p_{23}|\downarrow\rangle+p_{24}|\uparrow\downarrow\rangle$ for the wide and narrow bands, respectively.
The dependence of the probabilities $p_{ls}^{2}$ for the zero occupied state $|0\rangle$, the spin-up occupied state $|\uparrow\rangle$, the spin-down occupied state $|\downarrow\rangle$, and the double-occupied state $|\uparrow\downarrow\rangle$ in the ground state of the effective impurity model on $U$ are shown in Fig.~\ref{fig:pev}(a).
Corresponding to the three phases of metal, OSMP, and insulator, the ground states exhibit distinct characteristics when $J=U/4$ is considered:
(1) Within the metallic phase with $U\leq3.0$, the superposition states of the wide and narrow bands are both particle-hole symmetric, with $p_{l1}^{2}=p_{l4}^{2}$ and $p_{l2}^{2} = p_{l3}^{2}$.
(2) In the region of $3.0< U\leq3.8$, the OSMP presents specific superposition states, which have $p_{12}^{2}\neq p_{13}^{2}$, $p_{22}^{2}\neq p_{23}^{2}$, as well as very small probabilities $p_{11}^{2}=p_{14}^{2}\in[0.02,0.06]$ and almost zero values of $p_{21}^{2}= p_{24}^{2}$.
(3) In the insulating phase for $U>3.8$, there are two degenerate solutions with opposite spin occupancies. If one solution has $p_{l3}^{2}=1.0$ and $p_{l1}^{2}=p_{l2}^{2}=p_{l4}^{2}=0$, i.e., $|\Phi_{imp}^{ ol}\rangle=|\downarrow\rangle$, then the other solution should be $|\Phi_{ imp}^{ol}\rangle=|\uparrow\rangle$.

In OSMP with $J=U/4$, it is important to note that the ground states of two bands are specific superposition states, in which the probabilities of $|\uparrow\rangle$ and $|\downarrow\rangle$ are not equal in each band ($p_{l2}^{2}\neq p_{l3}^{2}$).
The probabilities of $|0\rangle$ and $|\uparrow\downarrow\rangle$ remain equal for all changes in the Coulomb interaction $U$.
In particular, $|\Phi_{imp}^{\rm o2}\rangle=p_{22}|\uparrow\rangle +p_{23}|\downarrow\rangle$, indicating that $|\uparrow\rangle$ and $|\downarrow\rangle$ form an entangled pair of qubits.
This state cannot be expressed as a product of states of its component systems \cite{Nielsen2010}. 
It is straightforward to demonstrate that $|\Phi_{imp}^{\rm o2}\rangle$ is an entangled state of the two-qubit system.
Assuming that $|\Phi_{imp}^{\rm o2}\rangle$ can be represented as the tensor product of two single quantum states,
$|\Phi_{up}^{\rm o2}\rangle=a_{2}|0\rangle +b_{2}|\uparrow\rangle$ and
$|\Phi_{down}^{\rm o2}\rangle=c_{2}|0\rangle +d_{2}|\downarrow\rangle$,
their tensor product is given by
$|\Phi_{up}^{\rm o2}\rangle |\Phi_{down}^{\rm o2}\rangle=a_{2}c_{2}|0\rangle+a_{2}d_{2}|\downarrow\rangle+b_{2}c_{2}|\uparrow\rangle+b_{2}d_{2}|\uparrow\downarrow\rangle$.
Therefore, it must hold that $a_{2}c_{2}=0$, $b_{2}d_{2}=0$, $a_{2}d_{2}\neq0$, and $b_{2}c_{2}\neq0$.
However, it is impossible to satisfy both $a_{2}c_{2}=0$ and $b_{2}d_{2}=0$ while also satisfying $|\Phi_{imp}^{\rm o2}\rangle$.
Hence, $|\Phi_{imp}^{\rm o2}\rangle$ is an entangled state.
These results illustrate the existence of intraorbital entanglement in the narrow band of OSMP with $J=U/4$.

Moreover, the results presented in Fig.~\ref{fig:pev}(a) additionally demonstrate that the sum rules $p_{12}^{2}+p_{23}^{2}\approx 1.0$ and $p_{13}^{2}+p_{22}^{2}+p_{11}^{2}+p_{14}^{2}\approx 1.0$ are valid.
With increasing $U$, the probabilities of wide band $|\downarrow\rangle$ and narrow band $|\uparrow\rangle$ both increase, while the probabilities of the wide band $|\uparrow\rangle$ and the narrow band $|\downarrow\rangle$ both decrease.
As a result, the probabilities of $|\uparrow\rangle$ and $|\downarrow\rangle$ in the two bands are no longer the same, suggesting the presence of quantum entanglement between the wide and narrow bands.
The intraorbital entanglement in narrow band and the interorbital entanglement between the two bands may collectively contribute to this finding.
The non-semi-integer values of LTQF in OSMP provide evidence of the presence of quantum entanglement, which is strongly influenced by Hund's coupling.
To this end, we calculated the case in the absence of Hund's coupling ($J=0$), as shown in Fig.~\ref{fig:jnb}(d). As we predicted, 
both $|\Phi_{\rm imp}^{\rm o1}\rangle$ and $|\Phi_{\rm imp}^{\rm o2}\rangle$ exhibit characteristics similar to the states with $J=U/4$, both in the metallic and insulating phases.
In OSMP with $J=0$, however, the wide band in the ground state can be described by an ordinary superposition state $|\Phi_{imp}^{\rm o1}\rangle=p_{11}|0\rangle+p_{12}|\uparrow\rangle +p_{13}|\downarrow\rangle+p_{14}|\uparrow\downarrow\rangle$, where spin-up and spin-down, zero and double occupied states always hold symmetry ($p_{12}^{2}=p_{13}^{2}$, $p_{11}^{2}= p_{14}^{2}$),
the narrow band is also no longer a specific superposition state but a double degenerate state, which can be expressed as
$|\Phi_{imp}^{\rm o2}\rangle=|\downarrow\rangle$ or $|\Phi_{imp}^{\rm o2}\rangle=|\uparrow\rangle$.
Apparently, the ordinary superposition state of each band in the two-band model with $J=0$ is more in accordance with the behavior of the one-band model \cite{Niu2023}.
To further prove that the non-semi-integer values of LTQF represent quantum entanglement, we focus on the entanglement entropy under the same conditions ($J=U/4$) in the next section.

\subsection{Hund's coupling driven quantum entanglement}
The entanglement entropy, as an auxiliary tool, is used to measure quantum
entanglement \cite{Chen2022,Dana2020,Lev2017}, which is helpful in identifying quantum phase transitions \cite{Korepin2004,Gu2004,Fernando2015,Fromholz2020}, such as characterizing
the Mott MIT \cite{Tiago2013,Tobias2013,Su2013,Walsh2019,Walsh2020,Canella2021}.
The von Neumann entropy, defined as $E_{v}=-{\rm Tr}(\rho {\rm log}\rho)$ \cite{Stephen2010},
has been utilized in the analysis of the half-filled one-band Hubbard model, also known as local entanglement  \cite{Gu2004}.
We analyze the local entanglement of the half-filled two-band Hubbard model by extending the quantity of the one-band Hubbard model as
\begin{equation}\label{EE}
E_{vl}=-2D_llog_{2}D_l-2(1/2-D_l)log_{2}(1/2-D_l),
\end{equation}
where $D_l=\langle n_{l\uparrow}n_{l\downarrow}\rangle$.
Here, the entanglement entropy depends on the correlation function of spin-up and spin-down electron occupations, which shows the correlation of local electronic states. 
The $|\Phi_{imp}^{ol}\rangle$ of LQSF is a superposition of four occupation states, belonging to a two-qubit state.
They all depend on the local electronic occupancy states.

\begin{figure}
\centering
\includegraphics[width=0.46\textwidth]{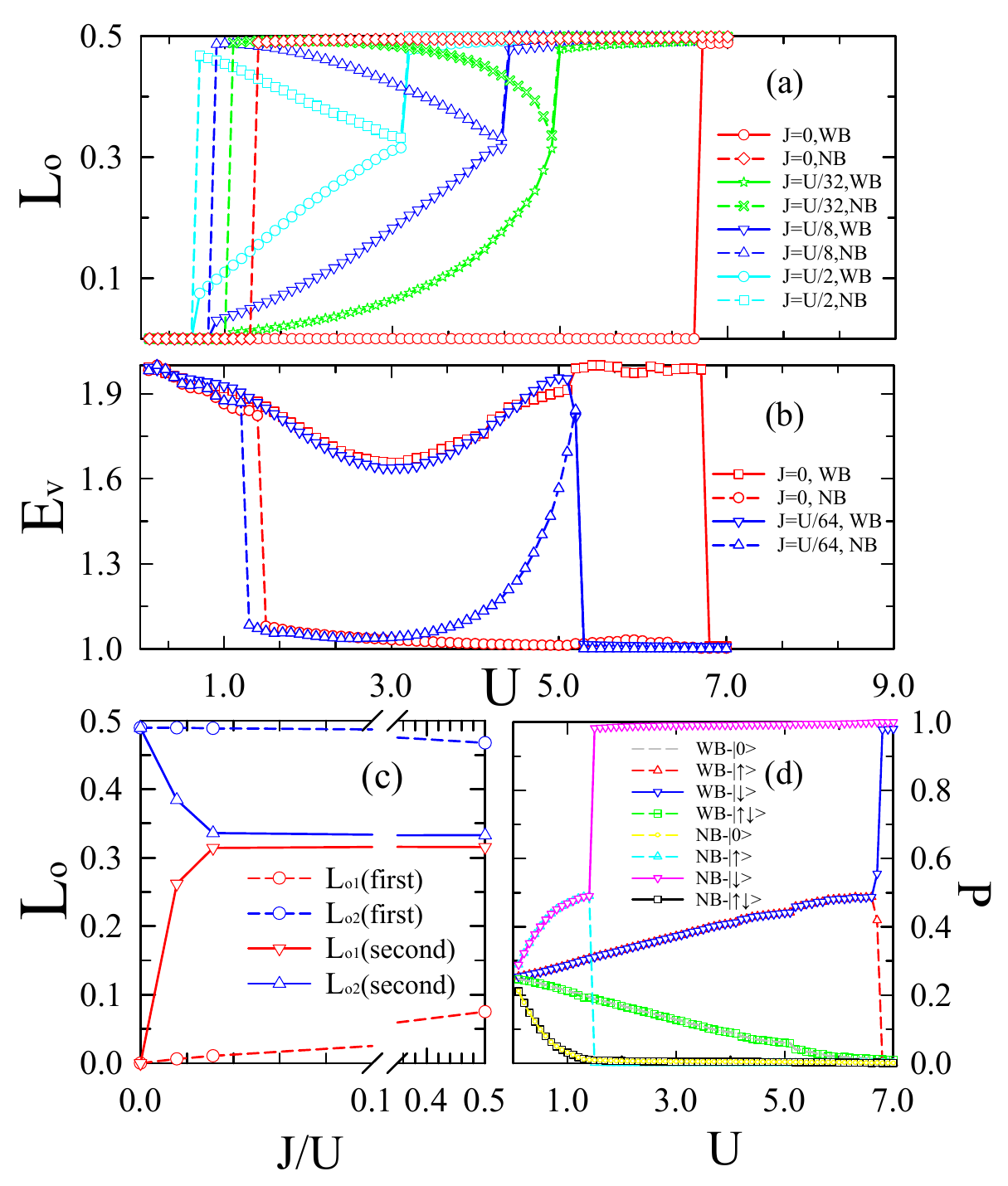}
\caption{\label{fig:jnb}(Color online)
(a) LTQF as a function of $U$ in conditions with different Hund's coupling strengths: $J=0$, $J=U/32$,
$J=U/8$, and $J=U/2$.
(b) The local entanglements $E_{v1}$ and $E_{v2}$ of the wide band and narrow band, respectively,
as a function of $U$ for two distinct scenarios: $J=0$ and $J=U/64$.
(c) The $J/U$ dependence of $L_{\rm o1}$ and $L_{\rm o2}$ at the first critical point (dashed line) and the second critical point (solid lines) within OSMP, respectively.
(d) The $U$ dependencies of the probabilities $p_{ls}^{2}$ of the empty state
$|0\rangle$, the spin-up state
$|\uparrow\rangle$, the spin-down state
$|\downarrow\rangle$, and the double-occupied state
$|\uparrow\downarrow\rangle$ at the single
impurity site for $J=0$.
The bandwidth ratio is set to $R=0.1$.}
\end{figure}

The local entanglement of the wide band $E_{v1}$ (shown as a pink solid line) and the narrow band $E_{v2}$ (shown as a cyan dotted line) as a function of $U$ is illustrated in Fig.~\ref{fig:pev}(b).
The model parameters used in this plot are $J=U/4$ and $R=0.6$. 
The local entanglement of each band has two cusps, which occur precisely at the two critical points obtained from LTQF shown in Fig.~\ref{fig:zl}(a).
It is predicted that the evolution of the local entanglement displays a sudden change at the critical point, which is caused by the energy level crossing of the ground state of a one-dimensional correlated fermion system \cite{Tian2003,Gu2004}.
Conversely, the local entanglement displays a sole cusp in each band as $U$ increases when $J=0$, as shown in Figs.~\ref{fig:pev}(b)(inset) and ~\ref{fig:jnb}(b).
In Fig.~\ref{fig:jnb}(b), the $E_{v2}$ exhibits the extreme value at the first critical point in OSMP, while $E_{v1}$ gradually rises before the second critical point in the non-hybridized two-band model.
However, the results of $J=U/4$ demonstrate that the local entanglement of each band exhibits two cusps
and the local entanglements of the two bands increase simultaneously with increasing $U$.
Therefore, it is proposed that the energy level crossing of the ground state leads to the interorbital entanglement, resulting in the correlation of $E_{v1}$ and $E_{v2}$ for a finite Hund's coupling [see Fig.~\ref{fig:pev}(b)].
In the absence of interorbital entanglement, $E_{v1}$ and $E_{v2}$ should be independent in OSMP [see Figs.~\ref{fig:pev}(b)(inset) and ~\ref{fig:jnb}(b)].
Quantum entanglement results in unequal probabilities for the states $|\uparrow\rangle$ and $|\downarrow\rangle$ within each band, leading to the non-semi-integer values of LTQF.
The above results indirectly confirm the existence of the interorbital entanglement in OSMP.

As previously stated, quantum entanglement can be identified by non-semi-integer values of LTQF in OSMP.
A defining characteristic of this phenomenon is the differentiation of superposition states based on Hund's coupling presence.
It is clear that physical mechanisms responsible for OSMT vary greatly depending on the presence or absence of Hund's coupling.
For instance, it is observed that OSMP does not exhibit quantum entanglement when $J=0$, whereas quantum entanglement is evident when $J\neq0$.
The influence of Hund's coupling on the behaviors of LTQF can be observed in Fig.~\ref{fig:jnb}(a),
where $L_{ol}$ as a function of $U$ at different Hund's coupling strengths $J=0$ (red line),
$J=U/32$ (green line), $J=U/8$ (blue line), and $J=U/2$ (cyan line) are presented for $R=0.1$.
The dashed and solid lines represent the values of LTQF of narrow band and wide band, respectively.
In the metallic phase, $L_{\rm o1}=0$ and $L_{\rm o2}=0$ for wide band and narrow band, respectively.
In the insulating phase, $L_{\rm o1}=0.5$ and $L_{\rm o2}=0.5$.
In particular, in OSMP with a finite $J$, $L_{\rm o1}$ increases and $L_{\rm o2}$ decreases with increasing $U$, and they intersect at the second critical point.
As $J$ increases, the ratios of change in LTQF exhibit varying monotonicity. 
The above results also prove that the non-semi-integer values of $L_{ol}$ in OSMP are a representation of the quantum entanglement.

The influence of Hund's coupling on LTQF of different bands is significant in OSMP.
In Fig.~\ref{fig:jnb}(c), we plot $L_{\rm o1}$ and $L_{\rm o2}$ as a function of $J$ at the first critical point (dashed line) and at the second critical point (solid line) of the narrow band (blue line) and wide band (red line), respectively.
As $J$ increases, $L_{\rm o1}$ exhibits a linear increase and $L_{\rm o2}$ decreases at the first critical point.
At the second critical point, $L_{\rm o1}$ increases nonlinearly for $J/U\lesssim\frac{1}{32}$, while $L_{\rm o2}$ decreases.
However, the values of $L_{\rm o1}$ and $L_{\rm o2}$ do not change
when $J/U\gtrsim\frac{1}{32}$, indicating that the Hund's coupling has a notable impact on OSMP.
In addition, Fig.~\ref{fig:jnb}(b) illustrates $E_{v1}$ and $E_{v2}$ as a function of $U$ in conditions with $J=0$ and $J=U/64$.
There is no correlation between $E_{v1}$ and $E_{v2}$ (red line) in OSMP when $J=0$, but they are correlated when $J=U/64$. 
This is the reason why the values of LTQF change faster near the second critical point.
It is worth noting that LTQF is no longer non-semi-integer in OSMP when $J=0$, specifically, $L_{o1}=0$ and $L_{o2}=0.5$ , as shown in Fig.~\ref{fig:jnb}(a).
This indicates that the quantum entanglement disappears for $J=0$.
Therefore, we illustrate the probabilities of electronic occupancy states at the single impurity site in Fig.~\ref{fig:jnb}(d).
The wide band is an ordinary superposition state, with $p_{11}^{2}=p_{14}^{2}$ and $p_{12}^{2} = p_{13}^{2}$. 
In contrast, the narrow band is a degenerate state, characterized by $p_{23}^{2}=1.0$.
It is suggested that the wave functions of each band represent ordinary superposition states that lack quantum entanglement for $J$=0.
The results depicted in Figs.~\ref{fig:jnb}(b) and ~\ref{fig:pev}(b) illustrate the coherent change in trend of the local entanglement within OSMP as $J$ increases.
This finding indicates a progressive enhancement of the quantum entanglement between the two bands as the $J$ is increased.

\subsection{The contributions of longitudinal and transverse Hund's couplings to the quantum entanglement}
The Hund's coupling can be decomposed into two components: the longitudinal and transverse terms.
The longitudinal term corresponds to the Ising-type Hund's coupling, while the transverse terms consist of the spin-flip and pair-hopping Hund's couplings \cite{Quan2015,Quan2017}.
To investigate the effects of different Hund's coupling terms on quantum entanglement, we examine
the interaction dependencies of LTQF in two distinct models. 
Specifically, we consider scenarios where the transverse Hund's coupling is absent and only the Ising term, denoted as the $J_z$ model.
When considering the full Hund's coupling terms, the corresponding model is the $J$ model.
This allows us to differentiate the influences of these coupling terms on quantum entanglement.
We find that the behaviors of LTQF are quite different with or without the $J_{sf}$ and $J_{ph}$.

\begin{figure}
\centering
\includegraphics[width=0.46\textwidth]{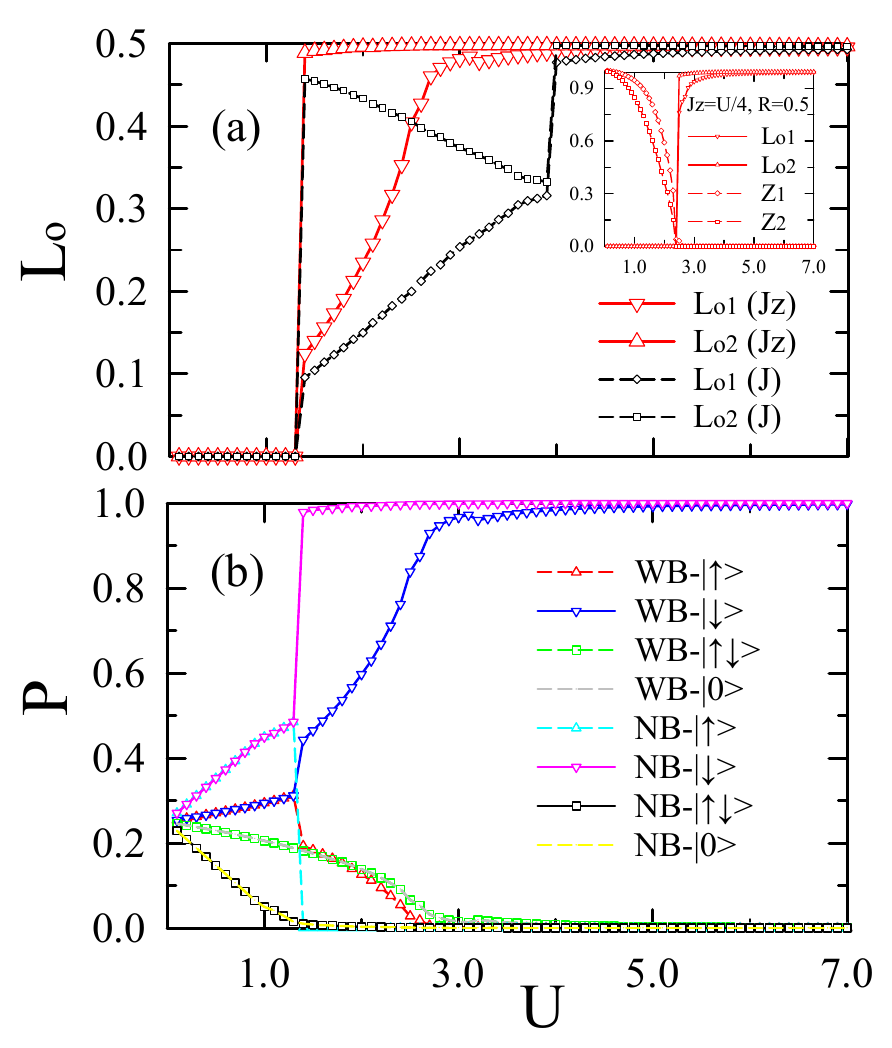}
\caption{\label{fig:jz}(Color online) (a) Comparing $L_{\rm o1}$ and $L_{\rm o2}$ of the two-band Hubbard models with or without the transverse Hund's coupling. For the transverse Hund's coupling, the corresponding critical points are $U_{c1}=2.7$ and $U_{c2}=1.4$. For the complete Hund's coupling, the corresponding critical points are $U_{c1}=4.0$ and $U_{c2}=1.4$ when $J=U/4$. 
(b) The probabilities $p_{ls}^{2}$ of the empty state $|0\rangle$, the spin-up state $|\uparrow\rangle$, the spin-down state $|\downarrow\rangle$, and double-occupied state $|\uparrow\downarrow\rangle$ at the single impurity site as a function of $U$.
The model parameters are $J_z=U/4$ and $R=0.2$.
(Inset) The evolution of the $L_{ol}$ and $Z_{l}$ with $U$ for $R=0.5$, $J_{z}=U/4$.}
\end{figure}

According to the result in Ref. \cite{Koga2005}, the quasiparticle coherent weights approach each other, yielding very close transition points for the $J_{z}$ model ($J_{sf}=J_{ph}=0$).
We calculated the results of quasiparticle coherent weights $Z_{l}$ and LTQF $L_{ol}$ under the same conditions and found that the quasiparticle coherent weights are indeed very close transition points, but LTQF can clearly distinguish the values at the transition point, as shown in Fig.~\ref{fig:jz}(a)(Inset).
This result once again demonstrates that using LTQF as a method to describe the Mott transition for studying anisotropy of the Hund's coupling is more accurate and convenient than using $Z_{l}$.
We also compare LTQF of the $J_z$ model with only the $J_{z}=U/4$ (red line) and $J$ model with the $J=U/4$ (black line), as shown in Fig.~\ref{fig:jz}(a). 
It is shown that the behavior of $L_{o2}$ of the $J_z$ model is the same as the case when $J=0$ shown in Fig.~\ref{fig:jnb}(a), where the $L_{o2}$ takes semi-integer values in OSMP.
Surprisingly, the behavior of $L_{o1}$ of the $J_z$ model is the same as the cases when $J\neq0$ in the $J$ model,
as depicted in Fig.~\ref{fig:jnb}(a), where $L_{o1}$ takes non-semi-integer values in OSMP.
A possible explanation for this might be the presence of "intraband entanglement" (meanwhile, with the presence of intraband unpaired doublons or holons \cite{Bittner2020,Yokoyama2022}) in the wide band for the $J_z$ model.
This question is left for a future investigation.

The probabilities $p_{ls}^{2}$ of the $J_z$ model with $R=0.2$ are shown in Fig.~\ref{fig:jz}(b).
We find that the probabilities of the narrow band satisfy $p_{21}^{2}=p_{24}^{2}=p_{22}^{2}=0$
(represented by the yellow, black and cyan lines) and $p_{23}^{2}=1.0$ (represented by the pink line),
while the wide band probabilities have $p_{11}^{2}=p_{14}^{2}\approx p_{12}^{2} \neq0$
(gray, green and red lines) and $p_{13}^{2}\neq1$ (blue line).
So, there is no interorbital entanglement in OSMP for the $J_z$ model.
This also indicates that the $J_{sf}$ and $J_{ph}$ play a crucial role in interorbital entanglement.
We consider that the transverse Hund's couplings lead to spin-antiparallel double occupied state, low-spin double occupied state \cite{Quan2017} and spin-orbital separation \cite{Deng2019}, which may promote the interorbital electron correlations in different orbitals.
Neglecting the $J_{sf}$ and $J_{ph}$ results in the disappearance of interorbital entanglement.
However, the $J_{z}$ can enhance the spin fluctuations of wide band .
Therefore, it is demonstrated that the physical mechanisms underlying the OSMT differ significantly depending on the presence or absence of the Hund’s coupling and its transverse terms.

\section{Conclusions\label{sec:cs}}
In summary, we have investigated the effects of Hund's coupling on quantum entanglement in the half-filled non-hybridized two-band Hubbard model using DMFT.
Similar to the one-band Hubbard model, MIT in the two-band Hubbard model can also be depicted by using LTQF, a valuable tool for exploring quantum entanglement.
Our research indicates that the occurrence of quantum entanglement in OSMP can be depicted by the non-semi-integer values of LTQF.
It is demonstrated that quantum entanglement is driven by Hund's coupling.
Specifically, the interorbital entanglement is mainly regulated by the transverse Hund's coupling, while the longitudinal term may generate intraorbital entanglement.
Meanwhile, the physical mechanisms underlying OSMT differ significantly depending on the presence or absence of Hund’s coupling and its transverse terms.
The specific differences can be summarized as follows:
({\romannumeral 1}) OSMP with $J = 0$ does not exhibit quantum entanglement, where the wide band is an ordinary superposition state and the narrow band is a single occupied state,
({\romannumeral 2}) OSMP with $J\neq 0$ ($J_{z}=J_{sf}=J_{ph}\neq 0$) exhibits quantum entanglement, where the wide band and narrow band are specific superposition states with the unequal probabilities of $|\uparrow\rangle$ and $|\downarrow\rangle$, 
and ({\romannumeral 3}) OSMP with $J_{z}\neq 0$ and $J_{sf}=J_{ph}=0$ exhibits "intraband entanglement" in wide band, where the wide band is a specific superposition states exhibiting the  unequal probabilities of $|\uparrow\rangle$ and $|\downarrow\rangle$, along with the equal probabilities of $|0\rangle$ and $|\uparrow\downarrow\rangle$.

LTQF can be easily used to handle two classes solution problems in multiorbital Hubbard model \cite{Liebsch2004}, which is also our next study to be carried out.
The present theory could be used to explain the anomalous properties in bad metals \cite{Emery1995,Deng2013,Mousatov2019,Pustogow2021}.
We also hope that it could promote the understanding of the non-Fermi-liquid state with frozen local moments \cite{Georges2013}.\\

\section*{Acknowledgments}

The authors would like to thank Professor Guang-Shan Tian and Associate Professor Lei Li for fruitful discussions.
Project supported by the Scientific Research Foundation for Youth Academic Talent of Inner Mongolia
University (Grant No.10000-23112101/010) and the Fundamental Research Funds for the Central Universities of China (Grant No. JN200208) and the Key Academic Discipline Project of China University of Mining and Technology (Grant No.2022WLXK11). 
Y.S. is supported by the National Natural Science Foundation of China (Grant No. 11474023). 
S.P. is supported by the National Key Research and Development Program of China (Grants No. 2023YFA1406500 and No. 2021YFA1401803) and the National Natural Science Foundation of China (Grant No. 12274036).

%

\end{document}